\documentclass[showpacs,preprintnumbers,10pt,twocolumn]{revtex4}%
\usepackage{amssymb}
\usepackage{amsfonts}
\usepackage{amsmath}
\usepackage{graphicx}
\usepackage{epsfig,float}
\usepackage{dcolumn}
\usepackage{bm}%
\begin{document}
\title{Impulse-Induced Optimum Signal Amplification in Scale-Free Networks}
\author{$^{1}$Pedro J. Mart\'{\i}nez and $^{2}$Ricardo Chac\'{o}n}
\affiliation{$^{1}$Departamento de F\'{\i}sica Aplicada, E.I.N.A., Universidad de Zaragoza,
E-50018 Zaragoza, Spain, and Instituto de Ciencia de Materiales de Arag\'{o}n,
CSIC-Universidad de Zaragoza, E-50009 Zaragoza, Spain}
\affiliation{$^{2}$Departamento de F\'{\i}sica Aplicada, E.I.I., Universidad de
Extremadura, Apartado Postal 382, E-06006 Badajoz, Spain, and Instituto de
Computaci\'{o}n Cient\'{\i}fica Avanzada, Universidad de Extremadura, E-06006
Badajoz, Spain}
\date{\today}

\begin{abstract}
Optimizing information transmission across a network is an essential task for
controlling and manipulating generic information-processing systems. Here, we
show how topological amplification effects in scale-free networks of signaling
devices are optimally enhanced when the \textit{impulse} transmitted by
periodic external signals (time integral over two consecutive zeros) is
maximum. This is demonstrated theoretically by means of a star-like network of
overdamped bistable systems subjected to \textit{generic} zero-mean periodic
signals, and confirmed numerically by simulations of scale-free networks of
such systems. Our results show that the enhancer effect of increasing values
of the signal's impulse is due to a correlative increase of the energy
transmitted by the periodic signals, while it is found to be resonant-like
with respect to the topology-induced amplification mechanism.

\end{abstract}

\maketitle

\textit{Introduction.}$-$Today, there is an emerging "network perspective"
approach to studying complex systems reflecting the ubiquitous presence of
networks in nature and in human societies. In particular, there has been
considerable interest in a class of networks known as scale-free networks due
to their lack of a characteristic size [1-4]. They have the property that the
degrees, $\kappa$, of the node follow a scale-free power-law distribution
$\left(  P\left(  \kappa\right)  \sim\kappa^{-\gamma},\gamma\in\lbrack
2,3]\right)  $. Examples are diverse metabolic and cellular networks, computer
networks such as the World Wide Web, and some social examples such as
collaboration networks. Besides topological investigations [5,6], current
interest in these (and other) networks has extended to their controllability
[7,8], i.e., to the characterization and control of the dynamical properties
of processes occurring in them, such as transport [9], synchronization of
individual dynamical behaviour occurring at a network's vertices [10,11], role
of quenched spatial disorder in the optimal path problem in weighted networks
[12], and dynamic pattern evolution [13]. One particular issue that has
attracted much interest because of its importance in both biological and
man-made information-processing systems is the propagation and enhancement of
resonant collective behaviour across a network due to the application of weak
external signals. In this regard, there has been recently studied the
amplification of the response to weak \textit{harmonic} signals in networks of
bistable signaling devices [14-17]. In these works, however, the robustness of
the signal amplification against \textit{diversity} in the uniform
distributions of periodic external signals was not studied. Clearly, the
assumption of harmonic external signals means that all driving systems$-$%
whatever they might be$-$are effectively taken as linear. This mathematically
convenient choice is untenable for most natural and artificial
information-processing systems due to their irreducible nonlinear nature.
Thus, to approach signal amplification phenomena in real-world networks, it
seems appropriate to consider distributions of periodic external signals which
are the output of \textit{nonlinear} systems, therefore being appropriately
represented by generic Fourier series.

In this Letter, we study the interplay between heterogeneous connectivity,
quenched spatial disorder, and generic zero-mean periodic signals in random
scale-free networks of signaling devices through the instance of a simple
deterministic overdamped bistable system. The system is given by
\begin{align}
\overset{.}{x}_{i}  &  =x_{i}-x_{i}^{3}+\tau f(t)-\lambda L_{ij}%
x_{j,}\;i=1,...,N,\nonumber\\
f(t)  &  \equiv\sum_{n=1}^{\infty}a_{n}\sin\left(  2n\pi t/T+\varphi
_{n}\right)  , \tag{1}%
\end{align}
where $f(t)$ is a unit-amplitude, zero-mean, $T$-periodic signal, $\tau$ is
the signal amplitude, $\lambda$ is the coupling, $L_{ij}=\kappa_{i}\delta
_{ij}-A_{ij}$ is the Laplacian matrix of the network, $\kappa_{i}=\sum
_{j}A_{ij}$ is the degree of node $i$, and $A_{ij}$ is the adjacency matrix
with entries 1 if $i$ is connected to $j$, and 0 otherwise. Since there are
infinitely many \textit{a priori }independent parameters, $a_{n}$,
$\varphi_{n}$, and hence infinitely many different \textit{waveforms} of
$f(t)$, the relevant problem is how to characterize quantitatively the effect
of the signal's waveform on the topology-induced amplification scenario [14].
Here, we shall show that a relevant quantity properly characterizing the
effectiveness of generic periodic signals (1) in the amplification scenario is
the \textit{impulse} transmitted by the signal over a half-period (hereafter
referred to simply as the impulse, $I\equiv\tau\int_{0}^{T/2}f(t)dt$) $-$ a
quantity integrating the conjoint effects of the signal's amplitude, period,
and waveform (see Fig. 1 for an example). Remarkably, we found that the
enhancer effect of increasing values of the impulse is due to a correlative
increase of the energy transmitted by the periodic signals. Extensive
numerical simulations of the system (1) were conducted for different network
topologies to characterize the effect of the impulse on the
amplification-synchronization scenario as the coupling strength is increased.
\begin{figure}[htb]
\epsfig{file=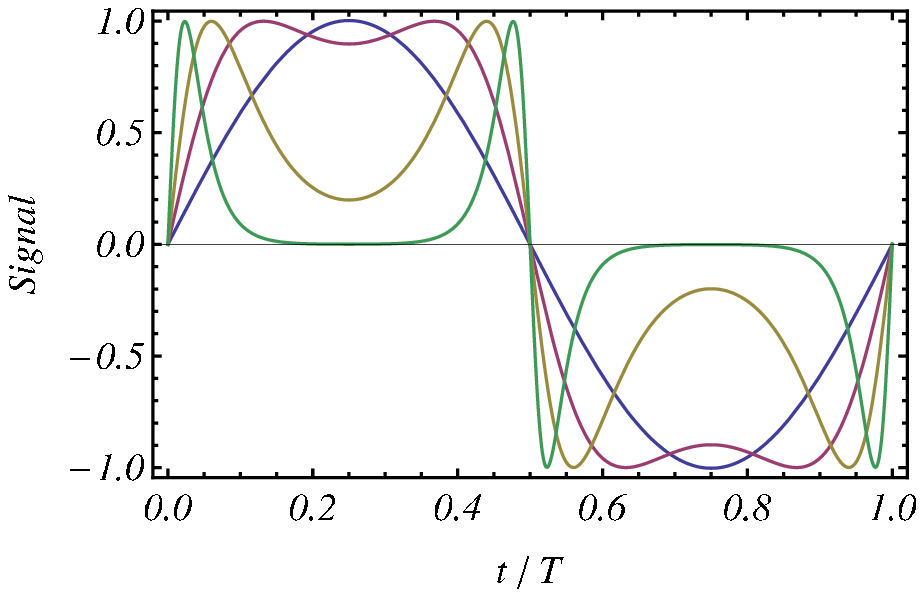,width=0.45\textwidth}
\epsfig{file=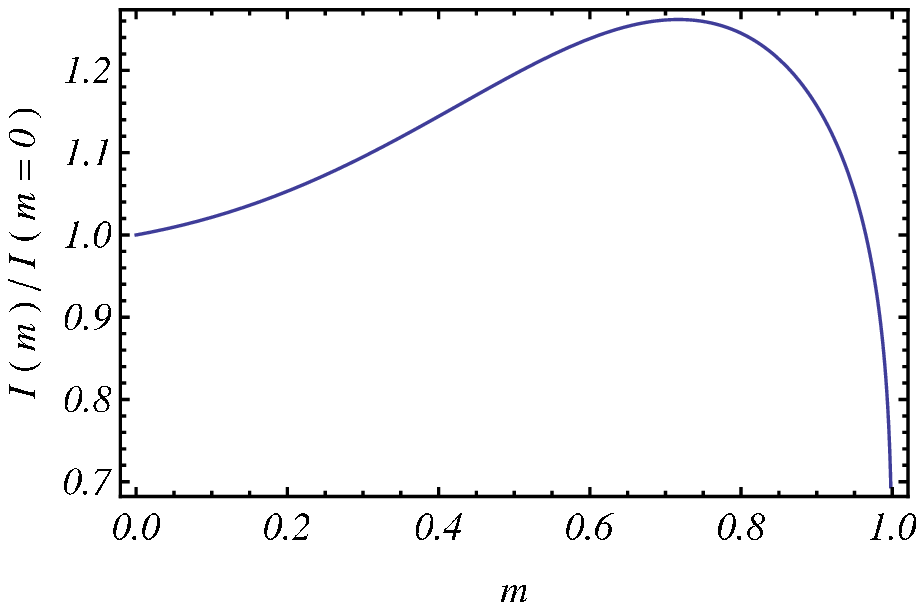,width=0.45\textwidth}
\caption{(Color online) Top: Signal $f\left(  t;T,m\right)  $ [Eq.~(12)] vs
$t/T$ where $T$ is the period and $N(m)\equiv1/\left\{  a+b/[1+\exp(\left\{
m-c\right\}  /d)]\right\}  $ with $a\equiv0.43932$, $b\equiv0.69796$,
$c\equiv0.3727$, $d\equiv0.26883$ for four values of the shape parameter:
$m=0$ (sinusoidal pulse), $m=0.72\simeq m_{\max}$ (nearly square-wave pulse),
$m=0.99$ (double-humped pulse), and $m=1-10^{-6}$ (sharp double-humped pulse).
Bottom: The normalized impulse $I(m,T)/I(0,T)$ [Eq.~(13)] vs $m$.}
\end{figure}
To quantitatively describe this scenario, we used the average amplification
$\left\langle G\right\rangle \equiv\max_{i}x_{i}/\tau$ over distinct initial
conditions on the one hand, and the synchronization coefficient [18]
\begin{equation}
\rho=\frac{\left\langle \overline{x_{i}}^{2}\right\rangle -\left\langle
\overline{x_{i}}\right\rangle ^{2}}{\overline{\left\langle x_{i}%
^{2}\right\rangle -\left\langle x_{i}\right\rangle ^{2}}}, \tag{2}%
\end{equation}
on the other, where the overlines indicate an average over nodes, while the
angle brackets indicate a temporal average over a period $T$.

\textit{Energy variation versus impulse}.$-$We start with a general argument
showing the relationship between energy increases and impulse increases in
isolated overdamped systems. Let us consider the family of dissipative
nonlinear oscillators $m\overset{..}{x}=-dU/dx-\delta\overset{.}{x}+\tau
f(t)$, with associated energy equation $\overset{.}{E}=\overset{.}{x}\left(
m\overset{..}{x}+dU/dx\right)  $, where $f(t)$ is a unit-amplitude, zero-mean,
$T$-periodic signal, $E(t)\equiv\left(  m/2\right)  \overset{.}{x}%
^{2}+U\left[  x\left(  t\right)  \right]  $ is the energy function with $U$
being a generic potential. The energy equation can be recast into the form%
\begin{equation}
\overset{.}{E}=-\delta\overset{.}{x}^{2}+\tau\overset{.}{x}f\left(  t\right)
. \tag{3}%
\end{equation}
Integration of Eq.~(3) over \textit{any} interval $\left[  nT,nT+T/2\right]
$, $n=0,1,2,...$, yields%
\begin{align}
E\left(  nT+T/2\right)   &  =E\left(  nT\right)  -\delta\int_{nT}%
^{nT+T/2}\overset{.}{x}^{2}\left(  t\right)  dt\nonumber\\
&  +\tau\int_{nT}^{nT+T/2}\overset{.}{x}\left(  t\right)  f(t)dt. \tag{4}%
\end{align}
Now, after applying the first mean value theorem for integration [19] to the
last integral on the right-hand side of Eq.~(4), one straightforwardly obtains%
\begin{equation}
E\left(  nT+T/2\right)  =E\left(  nT\right)  -\delta\int_{nT}^{nT+T/2}%
\overset{.}{x}^{2}\left(  t\right)  dt+\overset{.}{x}\left(  t^{\ast}\right)
I, \tag{5}%
\end{equation}
where $t^{\ast}\in\left[  nT,nT+T/2\right]  $ and $I$ is the impulse. Let us
consider an initial steady ($n$ large) situation fixing the parameters
$\left(  \delta,\tau,T\right)  $ and choosing a waveform such that the impulse
is relatively small. Next, we only change the waveform. For sufficiently small
values of $T$ and $\tau$, one expects that both the dissipation work (integral
in Eq.~(5)) and $\overset{.}{x}\left(  t^{\ast}\right)  $ will approximately
maintain their initial values, while $I$ may increase from its initial value
with the choice of a more convenient waveform, so that, in some cases
depending upon the remaining parameters, the energy difference $E\left(
nT+T/2\right)  -E\left(  nT\right)  $ will increase with respect to the
initial situation. Thus, the maximum probability of a maximal increase of the
energy difference occurs when $I$ is also maximum, which establishes a clear
correlation between impulse and energy transmitted over wide regions in
parameter space. Specifically, one expects this impulse principle to be
accurate when the signal period is the shortest significant timescale.
Numerical simulations confirmed the accuracy and scope of this prediction.
Figure 2 shows an illustrative example for an isolated overdamped bistable
system (cf. Eq. (1)).
\begin{figure}[htb]
\epsfig{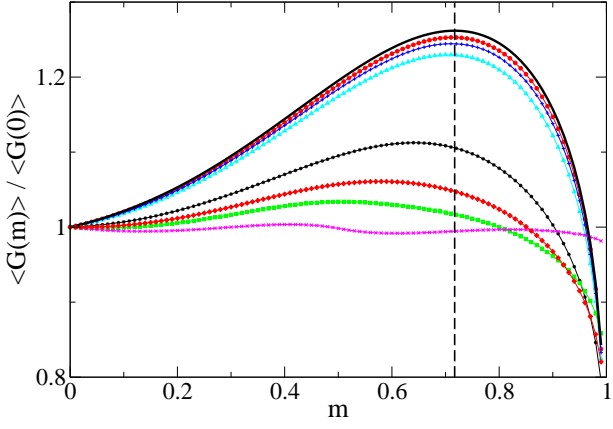}
\caption{(Color online) Normalized average amplification $\left\langle
G(m)\right\rangle /\left\langle G(m=0)\right\rangle $ versus shape parameter
$m$ for an isolated overdamped bistable system [Eq. (1) with $\lambda=0$]
subjected to a signal given by Eq. (12) and seven values of the signal period:
$T=1/\sqrt{2}$ (circles), $T=1\left(  +\right)  $, $T=\sqrt{2}$ (triangles),
$T=2\pi/\sqrt{2}\left(  \ast\right)  $, $T=7$ (diamonds), $T=10$ (squares),
and $T=50\left(  \times\right)  $. The solid line shows the normalized impulse
$I(m,T)/I(0,T)$ [Eq. (13)] versus shape parameter $m$ while the dashed
vertical line indicates the value $m=m_{\max}\simeq0.717$ for which
$I(m,T)/I(0,T)$ presents its single maximum as a function of $m$. Notice that
the period of the linearized motion around any of the potential minima,
$T=2\pi/\sqrt{2}$, corresponds to the shortest significant timescale.}
\end{figure}

\textit{Star-like network.}$-$Next, we consider a star-like network of
overdamped bistable systems:%
\begin{align}
\overset{.}{x}_{H}  &  =\left[  1-\lambda\left(  N-1\right)  \right]
x_{H}-x_{H}^{3}+\tau f(t)+\lambda\sum_{i=1}^{N-1}y_{i},\nonumber\\
\overset{.}{y}_{i}  &  =\left(  1-\lambda\right)  y_{i}-y_{i}^{3}+\tau
f(t)+\lambda x_{H}, \tag{6}%
\end{align}
which describes the dynamics of a highly connected node (or hub), $x_{H}$, and
$N-1$ linked systems (or leaves), $y_{i}$. We study the case of sufficiently
small coupling, $\lambda$, and external signal amplitude, $\tau$, such that
the dynamics of the leaves may both be decoupled from that of the hub and be
suitably described by linearizing their equations around one of the potential
minima. Thus, one straightforwardly obtains%
\begin{align}
y_{i}\left(  t\rightarrow\infty\right)   &  \sim\xi_{i}-\tau\sum_{n=1}%
^{\infty}\left[  C_{n}\cos\left(  \omega_{n}t\right)  +S_{n}\sin\left(
\omega_{n}t\right)  \right]  ,\nonumber\\
C_{n}  &  \equiv\frac{a_{n}\left(  \omega_{n}\cos\varphi_{n}-2\sin\varphi
_{n}\right)  }{\omega_{n}^{2}+4},\nonumber\\
S_{n}  &  \equiv-\frac{a_{n}\left(  \omega_{n}\sin\varphi_{n}+2\cos\varphi
_{n}\right)  }{\omega_{n}^{2}+4}, \tag{7}%
\end{align}
where $\omega_{n}\equiv2n\pi/T$ while $\xi_{i}=\pm1$ depending on the initial
conditions. Since the initial conditions are randomly chosen, this means that
the quantities $\xi_{i}$ behave as discrete random variables governed by
Rademacher distributions. After inserting Eq.~(7) into Eq.~(6) and solving the
resulting equation for the hub,
\begin{align}
\overset{.}{x}_{H}= & \left[  1-\lambda\left(  N-1\right)  \right]  x_{H}%
-x_{H}^{3}+\sum_{n=1}^{\infty}\left[  C_{n}^{\prime}\cos\left(  \omega
_{n}t\right) \right. \nonumber\\ 
& \left. +S_{n}^{\prime}\sin\left(  \omega_{n}t\right)  \right]
+\lambda\eta, \tag{8}%
\end{align}
where
\begin{align*}
\eta &  \equiv\sum_{i=1}^{N-1}\xi_{i},\\
\frac{C_{n}^{\prime}}{\tau}  &  \equiv a_{n}\left(  1+\frac{2\lambda\left(
N-1\right)  }{\omega_{n}^{2}+4}\right)  \sin\varphi_{n}-\frac{\lambda\left(
N-1\right)  a_{n}\omega_{n}}{\omega_{n}^{2}+4}\cos\varphi_{n},\\
\frac{S_{n}^{\prime}}{\tau}  &  \equiv\frac{\lambda\left(  N-1\right)
a_{n}\omega_{n}}{\omega_{n}^{2}+4}\sin\varphi_{n}+a_{n}\left(  1+\frac
{2\lambda\left(  N-1\right)  }{\omega_{n}^{2}+4}\right)  \cos\varphi_{n},
\end{align*}
one straightforwardly obtains%
\begin{align}
x_{H}\left(  t\rightarrow\infty\right) & \sim x_{H}^{(0)}+\sum_{n=1}^{\infty
}  \left[ \frac{\left(  C_{n}^{\prime}\omega_{n}+S_{n}^{\prime}a_{H}\right)
\sin\left(  \omega_{n}t\right)}{\omega_{n}^{2}+a_{H}^2} \right. \nonumber\\%
& \left. + \frac{\left(  C_{n}^{\prime}a_{H}-S_{n}^{\prime
}\omega_{n}\right)  \cos\left(
\omega_{n}t\right)}{\omega_{n}^{2}+a_{H}^2}\right], \tag{9}%
\end{align}
where $a_{H}\equiv V_{H}^{\prime\prime}\left(  x_{H}^{(0)}\right)  =-\left\{
\frac{3\lambda\eta}{x_{H}^{(0)}}+2\left[  1-\lambda\left(  N-1\right)
\right]  \right\}  $ with $x_{H}^{(0)}$ being the equilibrium in the absence
of any external signal, while $V_{H}(x_{H})\equiv-\sqrt{h}x_{H}^{2}+x_{H}%
^{4}/4$ is the hub potential with $h=\left[  1-\left(  N-1\right)
\lambda\right]  ^{2}/4$ being the height of the potential barrier. For finite
$N$, the quantity $\eta$ behaves as a discrete random variable governed by a
binomial distribution with zero mean and variance $N-1$. One sees that the
hub's dynamics is affected by spatial quenched disorder through the term
$\lambda\eta$, while an estimate of its amplification [20] is given by%

\begin{equation}
G\left[  f(t)\right]  =\frac{1}{\tau}\left(  \sum_{n=1}^{\infty}\frac
{C_{n}^{\prime2}+S_{n}^{\prime2}}{\omega_{n}^{2}+a_{H}^{2}}\right)  ^{1/2}.
\tag{10}%
\end{equation}
For sufficiently large $N$, we may assume that the quantity $\eta$ behaves as
a continuous random variable governed by a standard normal distribution, and
hence%
\begin{equation}
\left\langle G\right\rangle =\frac{1}{\sqrt{2\pi\left(  N-1\right)  }}%
\int_{-\infty}^{\infty}G\left[  f(t)\right]  \exp\left[  \frac{-\eta^{2}%
}{2\left(  N-1\right)  }\right]  d\eta\tag{11}%
\end{equation}
provides the final average amplification. Next, we demonstrate that the
impulse is the relevant quantity controlling the effect of the external signal
on the amplification by considering the illustrative example
\begin{equation}
f\left(  t;T,m\right)  \equiv N(m)\operatorname{sn}\left(  4Kt/T\right)
\operatorname*{dn}\left(  4Kt/T\right)  , \tag{12}%
\end{equation}
in which $\operatorname{sn}\left(  \cdot\right)  \equiv\operatorname{sn}%
\left(  \cdot;m\right)  $ and $\operatorname*{dn}\left(  \cdot\right)
\equiv\operatorname*{dn}\left(  \cdot;m\right)  $ are Jacobian elliptic
functions of parameter $m$ [$K\equiv K(m)$ is the complete elliptic integral
of the first kind] [21] and $N(m)$ is a normalization function (see Fig.~1,
top) which is introduced for the elliptic signal to have the same amplitude 1
and period $T$ for any wave form (i.e., $\forall m\in\left[  0,1\right]  $).
When $m=0$, then $f\left(  t;T,m=0\right)  =\sin\left(  2\pi t/T\right)  $,
i.e., one recovers the previously studied case of a harmonic signal [14],
whereas, for the limiting value $m=1$, the signal vanishes. Note that, as a
function of $m$, the impulse per unit of amplitude,%
\begin{equation}
I\mid_{\tau=1}=I(m,T)\equiv\frac{TN(m)}{2K(m)} \tag{13}%
\end{equation}
presents a single maximum at $m=m_{\max}\simeq0.717$ (see Fig.~1, bottom). In
this case, Eq.~(11) predicts that $\left\langle G\right\rangle \left(
\lambda,N,T,m=m_{\max}\right)  >\left\langle G\right\rangle \left(
\lambda,N,T,m\neq m_{\max}\right)  $ and that the signal amplification
\textit{increases }on average as the impulse is increased, i.e., as the shape
parameter $m\rightarrow m_{\max}$ (see Fig.~3, left), which is accurately
confirmed by numerical simulations (see Fig.~3, right). One also has from Eqs.
(11) and (12) that $\left\langle G\right\rangle \left(  \lambda,N,T,m\right)
$, as a function of only $\lambda$, presents a sharp single maximum at
$\lambda=\lambda_{\max}\simeq\left(  N-1\right)  ^{-1}$for all $m$, which
indicates that the topology-induced amplification mechanism is robust against
\textit{diversity} in the uniform distributions of periodic signals in
star-like networks.
\begin{figure}[htb]
\epsfig{file=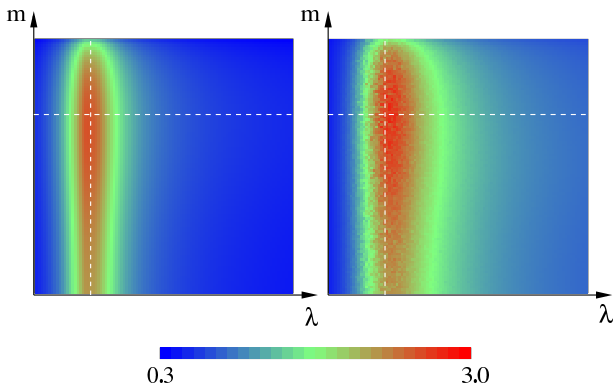,width=0.45\textwidth}
\caption{(Color online) Theoretical average amplification $\left\langle
G\right\rangle $ in the $\left(  \lambda-m\right)  $ parameter plane (left
panel, Eqs.~(11) and (12)) and corresponding numerical results $\left\langle
G\right\rangle $ (right panel) for a starlike network [Eq.~(6)] with
$\lambda\in\left[  0,0.01\right]  ,m\in\left[  0,0.99\right]  $ and
$N=500,T=10,\tau=0.01$. The dashed lines indicate the values $\lambda
=\lambda_{\max}\simeq0.002$, $m=m_{\max}\simeq0.717$ for which $\left\langle
G\right\rangle $ presents its single maximum in the $\left(  \lambda-m\right)
$ parameter plane.}
\end{figure}

\textit{Scale-free network.}$-$Next, we discuss the possibility of extending
the results obtained for a star-like network to Barab\'{a}si-Albert (BA)
networks [2] of the same overdamped bistable systems. Indeed, a highly
connected node in the BA network can be thought of as a hub of a local
star-like network with a certain degree $\kappa$ picked up from the degree
distribution. Thus, one can expect that the enhancer effect of the impulse
will act at any scale to yield a significant enhancement of the signal
amplification over the whole scale-free network in the weak coupling regime.
Figure 4 shows an illustrative example where the averaged amplification
$\left\langle G\right\rangle $ is plotted against the coupling $\lambda$ (top
panel) and the shape parameter $m$ (bottom panel). One sees that $\left\langle
G\right\rangle $ becomes ever larger as $m$ approaches $m_{\max}$ over the
complete range of values of $\lambda$, confirming the predictions of the above
theoretical analysis. Remarkably, the dependence of $\left\langle
G\right\rangle $ on $m$ follows in detail the respective dependence of the
impulse (see Figs.~1 bottom and 4 bottom). We found that this scenario remains
the same in any random realization of the network connectivity and for any
value of $m$ (see Fig.~5, top panel). The synchronization decreases
monotonically as $\lambda$ is increased for any value of $m$, with the
decrease being ever faster as $m\rightarrow m_{\max}$ (Fig.~5, bottom panel).
This can be understood as the result of two conjoint mechanisms: the
impulse-induced enhancement of amplification and the amplification-induced
lowering of synchronization in the weak coupling regime. Finally, we tested
the robustness of the present results against possible finite-size effects: we
found that the features of the impulse-induced amplification scenario remain
the same for much larger networks [22].
\begin{figure}[htb]
\epsfig{file=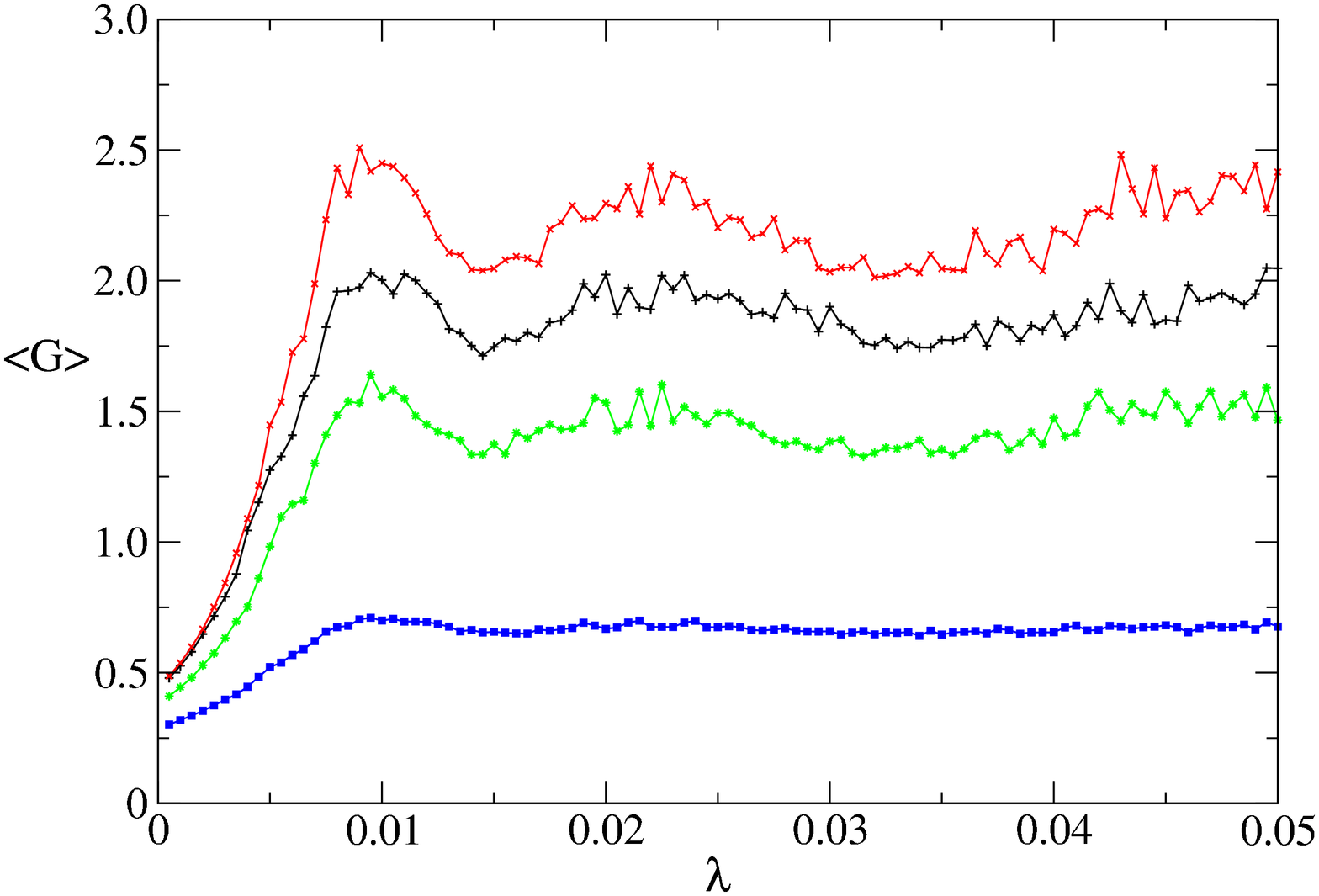,angle=-90,width=0.45\textwidth}
\epsfig{file=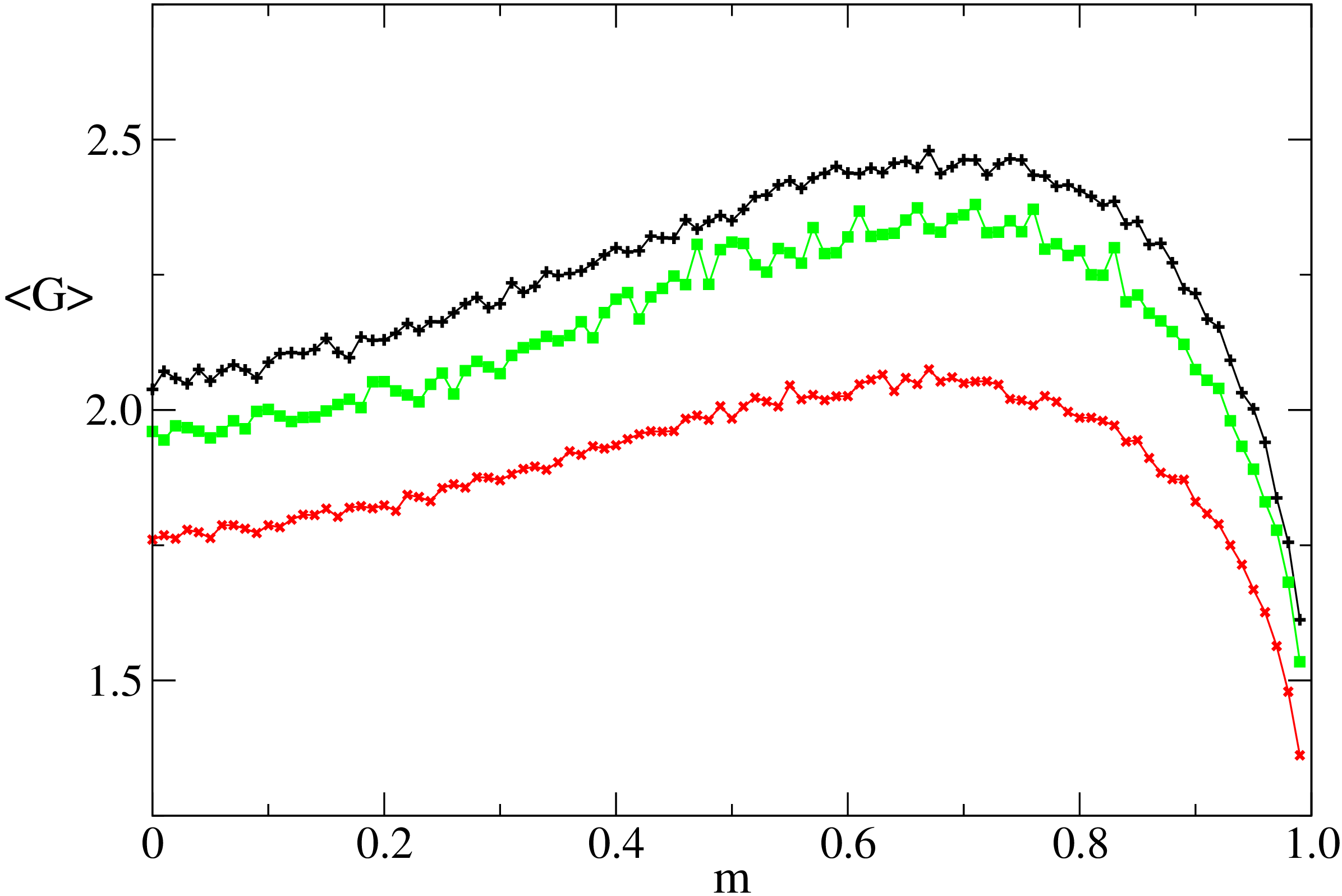,width=0.45\textwidth}
\caption{(Color online) Top panel: Average amplification $\left\langle
G\right\rangle $ versus coupling $\lambda$ for a BA scale-free network and
four values of the shape parameter: $m=0\ (+)$, $m=0.72\simeq m_{\max
}\ (\times)$, $m=0.99$ (stars), and $m=1-10^{-6}$ (squares). Note that the
first relative maximum of $\left\langle G\right\rangle $ occurs around
$\lambda\approx0.008$ for the four values of $m$ while the network has a
maximal active hub having $136$ leaves. For this effective star-like network
the theoretically predicted maximum occurs at $\lambda=\lambda_{\max,1}%
\approx0.0074$. Bottom panel: Average amplification $\left\langle
G\right\rangle $ versus shape parameter $m$ for a BA scale-free network and
three values of the coupling: $\lambda=0.009\ \left(  +\right)  $,
$\lambda=0.015\ (\times)$, and $\lambda=0.045$ (squares). Averaged degree
$\left\langle \kappa\right\rangle =3$, $\gamma=2.7$, and the remaining fixed
parameters are as in Fig.~3.}
\end{figure}
\begin{figure}[htb]
\epsfig{file=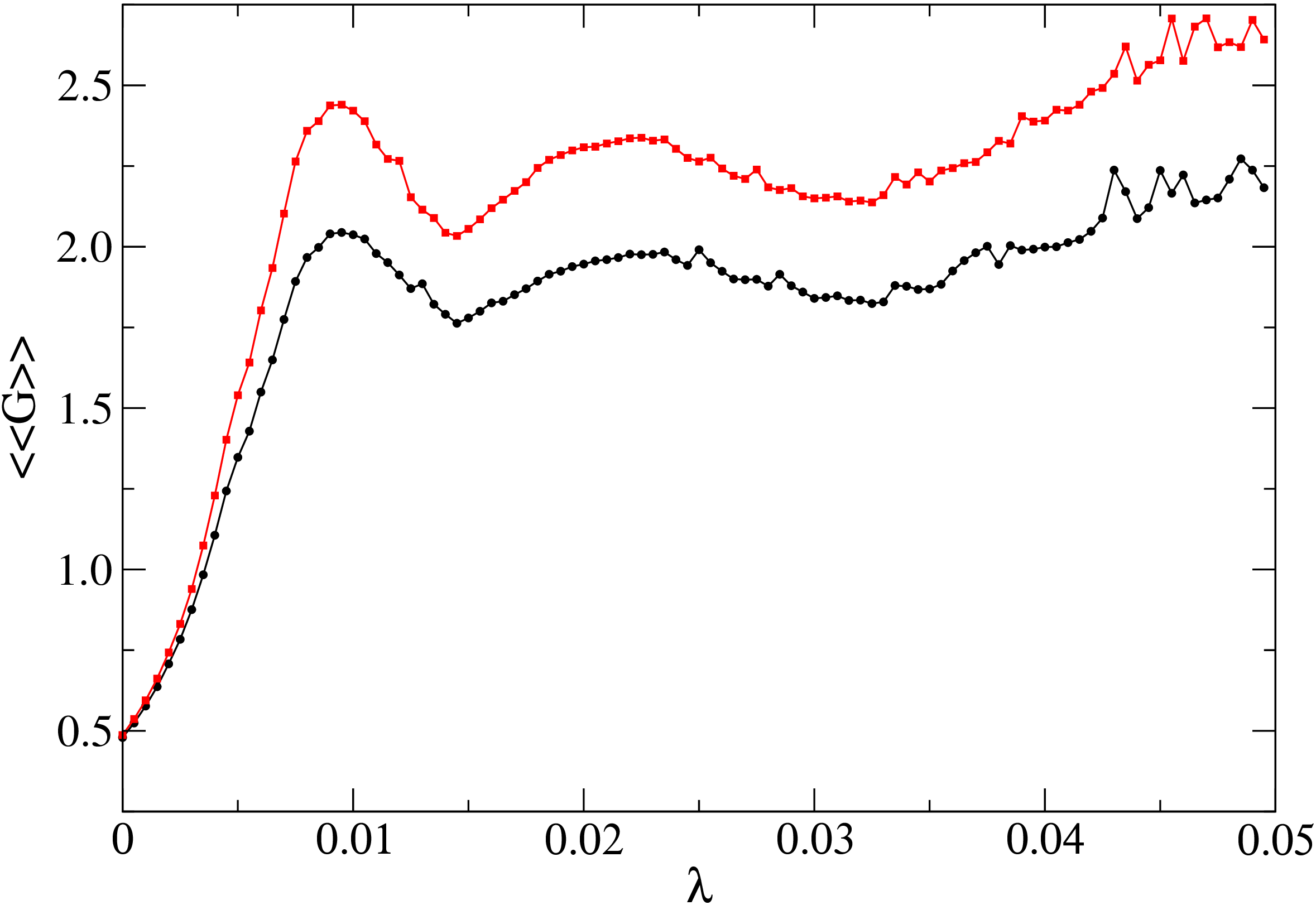,width=0.45\textwidth}
\epsfig{file=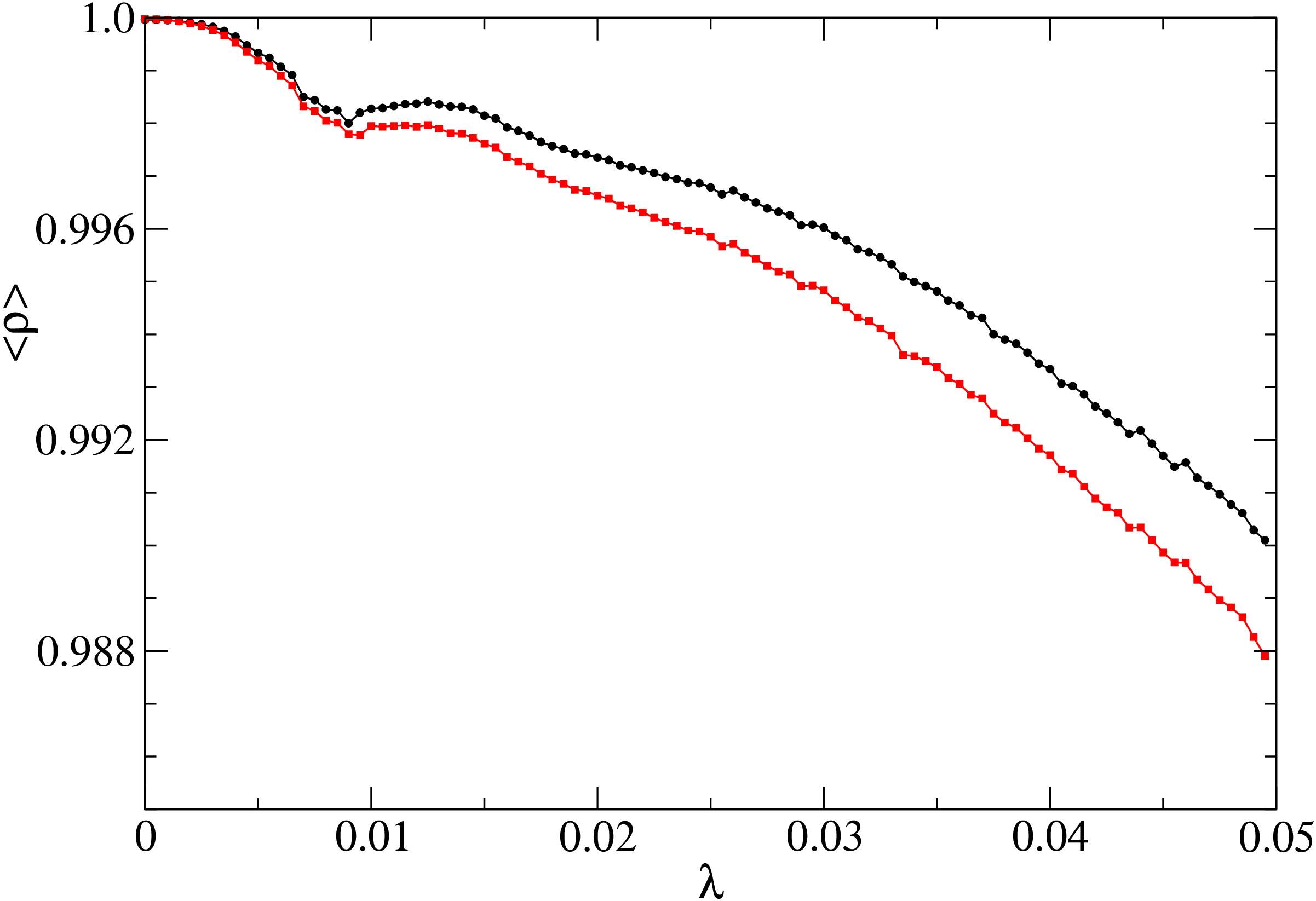,width=0.45\textwidth}
\caption{(Color online) Average of the average amplification over $10^{2}$
random realizations of the network connectivity $\left\langle \left\langle
G\right\rangle \right\rangle \equiv\left\langle \max_{i}x_{i}/\tau
\right\rangle $ for two values of the shape parameter ($m=0\ $(circles) and
$m=0.72\simeq m_{\max}\ $(squares)) (top panel) and corresponding
synchronization coefficient $\left\langle \rho\right\rangle $ [Eq.~(2)] for
$m=0$ and $m=0.72$ (bottom panel) versus coupling $\lambda$ for a BA
scale-free network with $\left\langle \kappa\right\rangle =3,\gamma=2.7$.
Other fixed parameters are as in Fig.~3.}
\end{figure}
\textit{Conclusion.}$-$We have shown through the example of a network of
overdamped bistable systems that maximizing the impulse transmitted by the
periodic external signals strongly enhances topology-induced signal
amplification in scale-free networks. We have analytically demonstrated that
this resonant-like effect of the impulse is due to a correlative increase of
the energy transmitted by the periodic signals as the impulse is increased,
while it may be completely characterized in the simple model of a starlike
network. Remarkably, our results indicate that varying the impulse does not
significantly change the values of the coupling strength for which
amplification is maximum, which means that the topology-induced amplification
mechanism is robust against diversity in the uniform distributions of periodic
external signals. Thus, the present findings provide a reliable criterion for
the optimization of topology-induced amplification processes in scale-free
networks by controlling the periodic external signals according to the impulse
principle. This novel principle opens up new avenues for studying
external-signal-induced amplification processes in complex networks,
including, for example, control of chaos by weak periodic excitations [23], or
optimization of signal transmission in neuronal networks [24,25] in which a
cooperative effect with the underlying noise of the neural medium is expected
under certain conditions.

P.J.M. and R.C. acknowledge financial support from the Ministerio de
Econom\'{\i}a y Competitividad (MINECO, Spain) through FIS2011-25167 and
FIS2012-34902 projects, respectively. P.J.M. acknowledges financial
support from the Comunidad de Aragón (DGA, Spain. Grupo FENOL) and European Social Funds.
 R.C. acknowledges financial support from
the Junta de Extremadura (JEx, Spain) through project GR15146.


\begin{thebibliography}{99}
\bibitem {20}This measure is based on the root-mean-square of a $T$-periodic function.
\end{thebibliography}
\end{document}